\documentclass[aps,prl,reprint,superscriptaddress]{revtex4-2}
\usepackage{graphicx}
\usepackage{dcolumn}
\usepackage{bm}
\usepackage[T1]{fontenc}
\usepackage{newtxtext,newtxmath}
\usepackage{amsmath}
\usepackage{nicefrac}
\usepackage{amsfonts}
\usepackage{microtype}
\usepackage{ulem}
\usepackage{color}
\usepackage{xcolor}
\usepackage[colorlinks=true,citecolor=blue,linkcolor=blue]{hyperref}

\newcommand{\fhi}{Molecular Physics, Fritz-Haber-Institut der Max-Planck-Gesellschaft, Faradayweg 4-6, 14195 Berlin, Germany}
\newcommand{\ikf}{Institut f\"ur Kernphysik, Goethe-Universität Frankfurt, Max-von-Laue-Straße 1, 60438 Frankfurt am Main, Germany}
\newcommand{\kassel}{Institut f\"ur Physik und CINSaT, Universität Kassel, Heinrich-Plett-Straße 40, 34132 Kassel, Germany}
\newcommand{\marburg}{Department of Chemistry, Philipps-Universität Marburg, Hans-Meerwein-Straße 4, 35032 Marburg, Germany}

\begin{document}
\title{Probing Instantaneous Single-Molecule Chirality in the Planar Ground State of Formic Acid}

\author{D.~Tsitsonis} \affiliation{\ikf}
\author{M.~Kircher} \affiliation{\ikf}
\author{N.~M.~Novikovskiy} \affiliation{\kassel}
\author{F.~Trinter} \affiliation{\fhi}
\author{J.~B.~Williams} \affiliation{Department of Physics, University of Nevada, Reno, Nevada 89557, USA}
\author{K.~Fehre} \affiliation{\ikf}
\author{L.~Kaiser} \affiliation{\ikf}
\author{S.~Eckart} \affiliation{\ikf}
\author{O.~Kreuz} \affiliation{\marburg}
\author{A.~Senftleben} \affiliation{\kassel}
\author{Ph.~V.~Demekhin} \email{demekhin@physik.uni-kassel.de} \affiliation{\kassel}
\author{R.~Berger} \email{robert.berger@uni-marburg.de} \affiliation{\marburg}
\author{T.~Jahnke} \email{jahnke@atom.uni-frankfurt.de}
\affiliation{Max-Planck-Institut für Kernphysik, Saupfercheckweg 1, 69117 Heidelberg, Germany}
\affiliation{European XFEL, Holzkoppel 4, 22869 Schenefeld, Germany}
\author{M.~S.~Sch\"offler} \email{schoeffler@atom.uni-frankfurt.de} \affiliation{\ikf}
\author{R.~D\"orner} \email{doerner@atom.uni-frankfurt.de} \affiliation{\ikf}

\date{\today}

\begin{abstract}
We experimentally demonstrate that individual molecules of formic acid are chiral even when they are in the vibronic ground state, which has a planar equilibrium structure. We ionize the C 1$s$ shell of the molecule and record the photoelectron in coincidence with positively charged fragments. This provides two consecutive measurements of the structure of one molecule, the first by photoelectron diffraction imaging and the second by Coulomb explosion imaging. We find that both measurements show the same handedness of the specific molecule. The phenomenon of being achiral on average but chiral at the level of individual molecules is general to most prochiral molecules and is a consequence of the three-dimensional zero-point delocalization of the nuclei in the vibrational ground state. 
\end{abstract}

\maketitle

Structures of rigid molecules are often represented using ball-and-stick models, which convey the classical perspective that atomic nuclei occupy fixed positions with well-defined arrangements, distances, and angles. However, even within the Born-Oppenheimer approximation, the vibrational wavefunction possesses a finite spatial extent, challenging the notion of nuclei as particles with precise locations. The quantum nature of atomic nuclei implies that if their positions in isolated molecules were measured, one would find a distribution of positions with probabilities given by the square of the vibrational wavefunction. For molecules with a planar equilibrium structure, this quantum feature has an important consequence: essentially any individual measurement will reveal a three-dimensional rather than a two-dimensional molecular arrangement, as the nuclei will be found slightly out of the equilibrium plane.
Extruding an object from two to three spatial dimensions can lead to the emergence of a new property absent in planar objects embedded in three-dimensional space: chirality. A notable molecular example is formic acid (HCOOH), which has a planar equilibrium structure in the electronic ground state. Due to quantum delocalization and zero-point motion of the nuclei, any measurement of an individual formic acid molecule will find a chiral structure; with R and S enantiomers appearing, neglecting the effect of parity violation due to the fundamental weak interaction \cite{QuackAngewCh1989,berger:2019}, with equal probability.

In the following, we choose the term ``instantaneous chirality'' for this quantum effect (see Refs.~\cite{HarrisChemPhy,Kitamura2001,MannChemPhy2014}). The established terms ``dynamical chirality'' \cite{QuackAngewCh1989,MarquardtPhysChA2000,Kitamura2001}  and ``transient chirality'' \cite{KlunderChemRev} do not quite capture the quantum nature of this phenomenon, as it arises in the stationary ground state of the molecule where there is no time dependence and no motion, and thus, no dynamics. Other than for transient chirality, the molecule is also not excited to an electronic state where it pyramidalizes out of its formerly planar shape and vibrates through a transiently chiral configuration \cite{Rouxel}.  It is one of the puzzling aspects of quantum theory that for a stationary state, instantaneous chirality does not already exist as a realistic property of the prochiral molecule before it is measured (see Ref.~\cite{FortinFoundChem2021}). It first emerges due to spontaneous symmetry breaking during the measurement process. For simplicity and readability, this Letter will, however, adopt the not entirely accurate ``realistic'' terminology, setting aside these complex quantum theoretical nuances.

The purpose of the present Letter is to report on an experiment in which we measure the handedness of individual formic acid molecules in the gas phase twice, in two consecutive measurements on the same molecule, exploiting inner-shell photoionization. The first measurement is done by photoelectron diffraction and the second one follows shortly after by imaging the Coulomb explosion as the molecule rapidly fragments. In photoelectron diffraction, a photoelectron wave is emitted from a localized $K$-shell ``illuminating the molecule from within'' \cite{Landers2001}. This creates a diffraction pattern in the molecular frame of reference. The power of this technique for structure determination has been shown in many works (see Ref.~\cite{JahnkeRoySocChem2023} for a review). 

Historically, the Coulomb explosion was first ignited by passing swift molecular ions through a thin foil, consequently stripping many electrons \cite{VagerScience1989}. Today, mostly neutral molecules are multiply charged by a strong femtosecond laser pulse \cite{PhysRevLett.74.3780,Schouder2022}, a photon pulse from an FEL \cite{NatureBoll.10.1038}, or by single-photon ionization of an inner-shell electron followed by Auger cascades \cite{muramatsu2002,Kilian2022NewRoute,KilianPECDtrifluoromethyloxirane,KilianPhysRevLett.126.083201}. In Coulomb explosion imaging, one measures the linear momentum vectors of ionic fragments that carry the desired information on the structure in coordinate space. In cases where the ionic potential surface is known exactly \cite{PhysRevA.102.063125}, with some constraints, even an inversion from the measured linear momenta into the initial coordinate space is possible \cite{PhysRevA.71.013415,PhysRevLett.108.073202,KunitskiScience2015,sayler2018}. The technique is well-suited for measuring deviations from the equilibrium structure. Examples include the precise determination of the helium dimer bond length \cite{Zeller2016} and the helium trimer geometry, where the notion of a well-defined triangular structure was shown to fail completely \cite{ZellerNatComm2014}. The sensitivity of this technique to ground-state fluctuations, i.e., to correlations between positions of atoms in the ground state of larger molecules, has recently been shown in Ref. \cite{Tillground}.

\begin{figure}
\center
\includegraphics[width=1.\columnwidth]{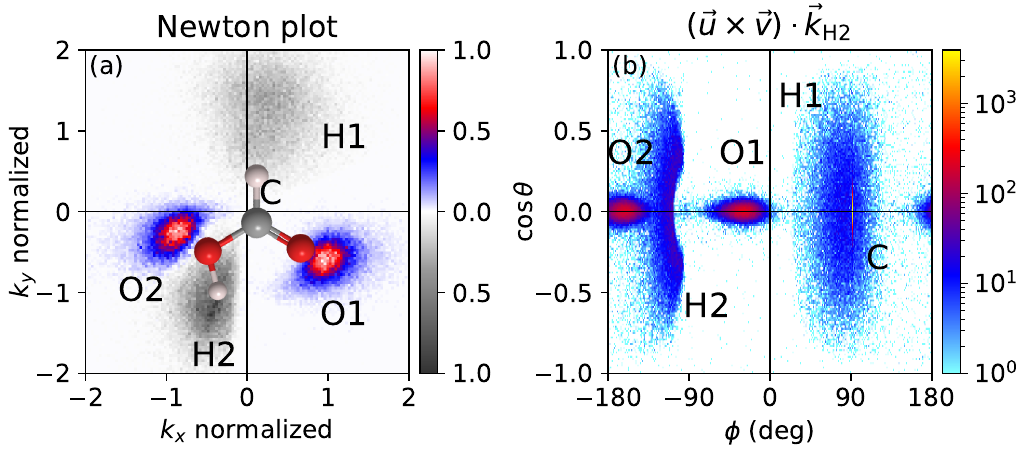}
\caption{Coulomb explosion imaging of formic acid following C~1s photoionization at a photon energy of 305.5~eV. Four charged particles (C$^+$, O$^+$, O$^+$, and H$^+$) are detected, and the linear momentum vector of the fifth particle (H) is obtained using linear momentum conservation. Panel~(a) shows the distribution of momenta in the plane of the molecule, the carbon ion is emitted to the right (see text for definition of coordinates). The magnitudes of the momenta of the particles $i$ are scaled by $1/\sqrt{m_i}$, where $m_i$ is the mass of each fragment. The color bar represents the normalized relative intensity of the data. Superimposed in this plot is the equilibrium structure of formic acid with C in grey, O in red, and H in white and a double bond being indicated between C and the carbonyl oxygen O1. (b) Polar representation of a side view onto the plane shown in panel (a). The plane is located at $\cos\theta=0$, where $\theta$ is the polar angle with respect to the normal of the plane in panel (a). $\phi=90^{\circ}$ is selected to be the direction of the carbon ion.}
\label{Fig1}
\end{figure}

The experiment was performed at beamline P04 of the synchrotron radiation facility PETRA~III at DESY (Hamburg, Germany) in 40-bunch mode \cite{Viefhaus2013}. The pulsed photon beam was crossed with a supersonic molecular beam in a COLTRIMS reaction microscope \cite{DORNER200095,Ullrich2003,JAHNKE2004229}, such that on average less than one molecule Coulomb-explodes per photon pulse. Electrons and ions were guided by an electric field of 160~V/cm onto microchannel-plate detectors with hexagonal delay-line position readout \cite{Jagutzki2002}. From the measured times-of-flight and positions-of-impact of the ions, the mass-over-charge ratios and the three-dimensional momentum vectors were calculated. To distinguish the two O and two H atoms, we followed the procedure described in the methods section of Ref.~\cite{Fehre2019}.

Figure~\ref{Fig1} shows the measured linear momentum vectors for the case where we detected four charged particles: C$^+$, O$^+$, O$^+$, and H$^+$. The momentum vector of the undetected H was obtained by exploiting linear momentum conservation. We find that the momentum vectors of the three heavy particles are almost coplanar, and we use them to define a coordinate frame where the unit vectors $\hat{x},\hat{y},\hat{z}$ are defined as follows:
\begin{equation}
\begin{aligned}
    \hat{z} &= \vec{u} \times \vec{v} / |\vec{u} \times \vec{v}| \ , \\
    \hat{x} &= \hat{z} \times \vec{w}/|\hat{z}  \times \vec{w}  | \ , \\
    \hat{y} &= \hat{z} \times \hat{x} \ , \\
\end{aligned}
\label{equ1}
\end{equation}
with
\begin{equation} 
\begin{aligned}
    \vec{u} &= \vec{k}_{\text{O2}} -\vec{k}_{\text{C}} \  , \\
    \vec{v} &= \vec{k}_{\text{O1}} -\vec{k}_{\text{C}} \ , \\
    \vec{w} &= \vec{k}_{\text{C}} \times (\vec{u} \times \vec{v}) \ .
\end{aligned}
\label{equ2}
\end{equation}
In this coordinate frame, the momenta of the heavy particles span the $x$-$y$ plane. Figure~\ref{Fig1}(a) shows a Newton-plot-type top view onto this molecular plane with the carbon ion momentum pointing to the right along the positive $x$ direction. The magnitude of the normalized momenta of particle $i$ shown in Fig.~\ref{Fig1}(a) has first been scaled by $1/\sqrt{m_i}$ (with $m_i$ being the mass of the respective particle) to account for the mass differences, and has then been normalized to the carbon $k_{\text{norm},i}=\nicefrac{(k_i/\sqrt{m_i})}{(k_{C}/\sqrt{m_c})}$. In Fig.~\ref{Fig1}(b), we show a ``side view'' onto the $x$-$y$ plane choosing polar coordinates with respect to the $z$ axis. Thus, the polar angle $\theta$ is the angle to the normal with respect to the $x$-$y$ plane shown in panel (a). The  $x$-$y$ plane is located at $\cos{\theta}=0$. The carbon ion momentum defines $\phi=90^{\circ}$. This representation, indeed, shows the linear momenta of the heavy nuclei to be narrowly confined to the plane at $\cos{\theta}=0$. The momenta of the two protons are bending broadly out of the molecular plane. The angles shown in the figure are angles between the measured linear momentum vectors, and the mapping from angles in position space to momentum space is highly nonlinear. Coulomb explosion, as a tool for sensing this out-of-plane extrusion of light particles, profits from a magnification upon changing from coordinate to final-state momenta (see, e.g., Supplementary Material of Refs.~\cite{Fehre2019} and \cite{Tillground}). To give an example, a wagging of H1 of 6$^\circ$ out of the plane results in an out-of-plane angle of 15$^\circ$ for the H1 linear momentum after the Coulomb explosion.  

\begin{figure}
\center
\includegraphics[width=1.\columnwidth]{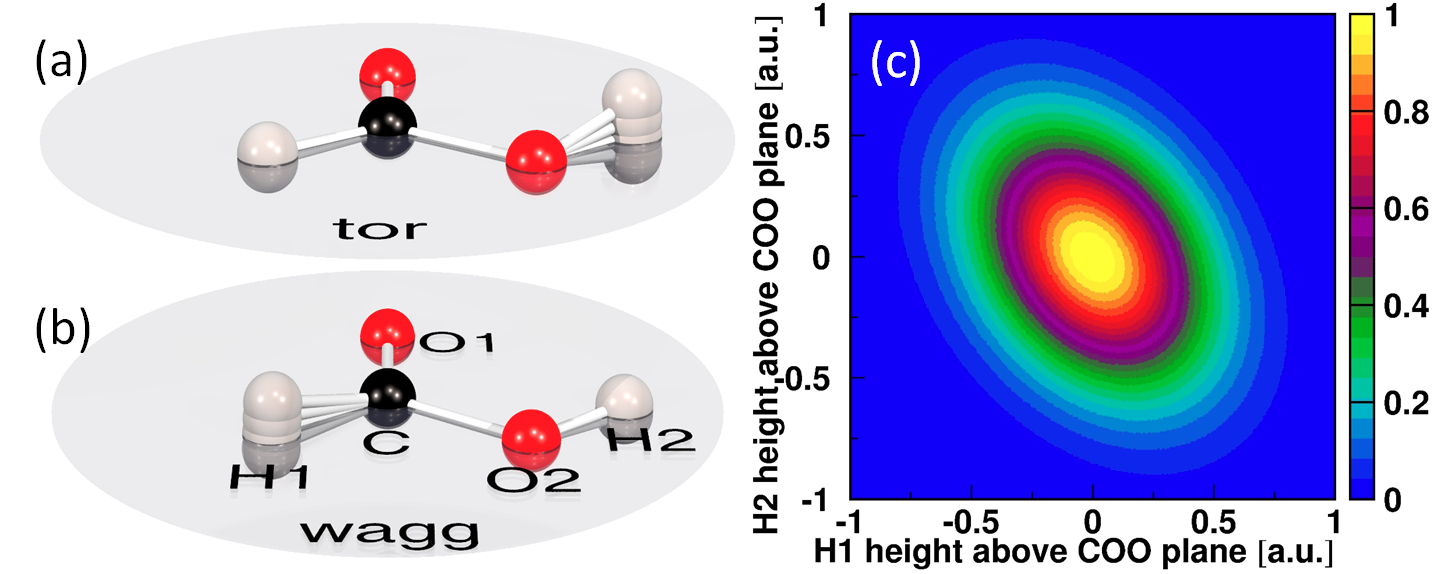}
\caption{Formic acid vibronic ground state. Panels~(a) and (b): structures between the classical turning points of the two out-of-plane normal modes of the formic acid electronic ground state as computed in the harmonic approximation on the explicitly correlated coupled-cluster levels with iterative singles, doubles, and perturbative triples amplitudes [CCSD(T)-F12] with a triple-zeta basis set (aug-cc-pVTZ-F12) using MOLPRO \cite{MOLPRO_brief,MOLPRO-WIREs}. Panel~(a): OH torsional mode (669.67~cm$^{-1}$). Panel~(b): CH wagging mode (1054.68~cm$^{-1}$). Panel~(c): computed correlation between the two H atoms in the vibrational ground state based on the reduced two-mode harmonic model. The horizontal and vertical axis show the height of H1 and H2 above the COO plane, respectively.}
\label{Fig2}
\end{figure}

In Fig.~\ref{Fig2}, we show the structure of the vibronic ground state of formic acid. There are nine vibrational normal modes: seven in-plane and two out-of-plane deformations. When considering small amplitude displacements in the vibrational ground state of formic acid, only the latter two are of interest in the current context, while for larger displacements, nonlinearities will kick in. Panels~(a) and (b) show the molecule's structure at the classical turning points of the respective torsional and wagging normal modes. In the harmonic approximation, the ground-state wavefunction as a function of the generalized coordinates of the normal modes is a product of two Gaussian functions. Coulomb explosion does not sense the normal-mode coordinates but the position coordinates of the individual atoms. The transformation of the ground state to this experimentally accessible coordinates is shown in Fig.~\ref{Fig2}(c). This shows that, within the ground state, the position of the two H atoms with respect to the plane spanned by the heavy atoms is highly anti-correlated. Such ground-state correlations have recently been experimentally confirmed, as well, for a larger molecule \cite{Tillground}.

\begin{figure}
\center
\includegraphics[width=1.\columnwidth]{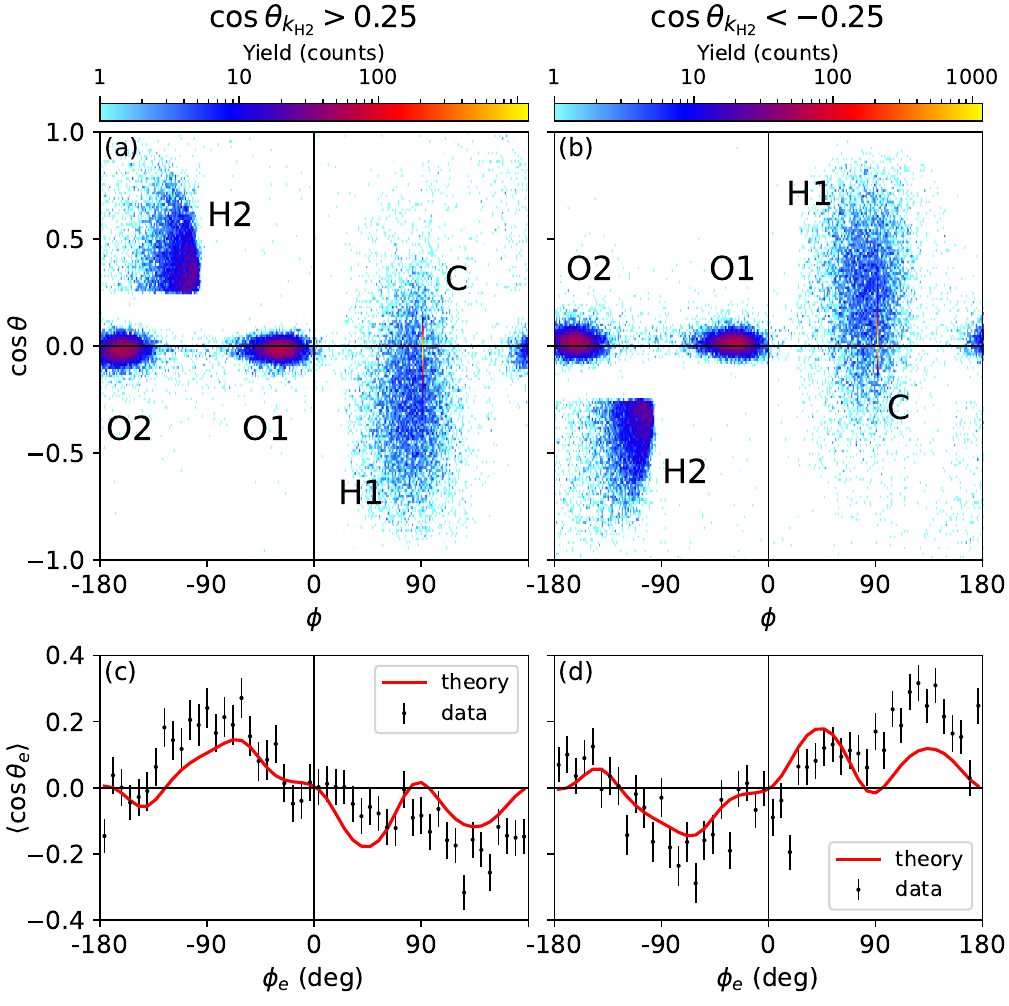}
\caption{Combined Coulomb explosion and photoelectron diffraction imaging of instantaneous chirality. (a,b) Side view onto the molecular frame as in Fig.~\ref{Fig1}(b). Panel (a) [Panel (b)] shows a subset of data, where the H2 atom is emitted to upward [downward], $\cos{\theta_{k_\mathrm{H2}}}>0.25$ (R enantiomer) [$\cos{\theta_{k_\mathrm{H2}}}<-0.25$ (S enantiomer)]. (c,d) Photoelectron diffraction imaging of the molecules selected as in panels (a) and (b). The vertical axis shows the mean out-of-plane angle of the photoelectrons. Histogram: experimental data; full lines: result of single-center calculation after convolving with an estimated experimental resolution of $20^{\circ}$. For the experimental data in panels (c) and (d), the three-body breakup channel H2$^+$, C$^+$, and O2$^+$ was used.}
\label{Fig3}
\end{figure}

The broad distribution of the out-of-plane angle of the proton momenta suggests the use of $\cos{\theta_{k_\mathrm{H2}}}$ as an indicator of the chirality. It is given by the triple product (see Refs. \cite{Pitzer_2016,Fehre2019})
\begin{equation}
\cos{\theta_{k_\mathrm{H2}}} = \frac{( \vec{u} \times \vec{v})\cdot \vec{k}_{\text{H2}}}{|\vec{u}\times\vec{v}| \cdot  |\vec{k}_{\text{H2}}|}
\label{equ3}
\end{equation}
Here, $\cos{\theta_{k_\mathrm{H2}}}>0$ corresponds to the R enantiomer and $\cos{\theta_{k_\mathrm{H2}}}<0$ to the S enantiomer. This allows unambiguously to split detected molecular fragmentation events into left- or right-handed enantiomers, even though they all arise from the planar ground state of formic acid. In Fig.~\ref{Fig3}, we select two subsets from our data corresponding to R [panels~(a) and (c)] and S [panels~(b) and (d)] enantiomers, respectively, choosing $\cos{\theta_{k_\mathrm{H2}}}>0.25~ (<-0.25)$ [i.e., H2 being emitted upward/downward in Fig~\ref{Fig1}(b)]. Figures~\ref{Fig3}(a) and \ref{Fig3}(b) show an anti-correlated wagging of the H1 fragment. This is caused, in part, by the anticorrelation in coordinate space (Fig.~\ref{Fig2}), and to a larger extent by the choice of the coordinates, as the $\vec{u}$ and $\vec{v}$ vectors (Eq.~\ref{equ2}) include the recoil of the hydrogen atoms by linear momentum conservation. Furthermore, anti-correlation between the two protons' linear momenta is also built up during the fragmentation by the Coulomb repulsion.

The chirality in the momentum pattern that we have discussed so far is measured after fragmentation. While we have argued that the observed \textit{momentum} asymmetry reflects symmetry breaking in the \textit{spatial} structure of the initial ground state, this is not the only possible explanation. Previous studies of multiple ionization of formic acid in a strong laser pulse concluded the opposite. They argued that symmetry breaking might occur during the dynamics on a potential energy surface of an electronically excited state after the promotion of an electron to an unoccupied $\pi^\ast$ orbital. This excited state is known to have a double-well potential with minima for R and S pyramidalized geometries \cite{Fridh,NgBell,MouleLimeEtAl}.

Along this question, for single-photon excitation, it has recently been shown and explained, in detail, how the light polarization can be used to drive this excited state preferentially into the R or S geometry \cite{Tsitsonis24prl}.
However, an alternative scenario---dynamical symmetry breaking on an excited-state surface---must be excluded. To do so,
we now turn to the angular emission pattern of the photoelectrons.
Photoelectron diffraction takes a snapshot of the molecular structure at time $t=0$, the instant of photoabsorption. It captures the structure of the initial molecular arrangement, before the nuclei could move and before the fragmentation has set in. The time between this snapshot and the start of the fragmentation is given by the Auger lifetime of the carbon $K$-shell hole, which is of the order of 7~fs \cite{Hergenhahn_2004}. 
To avoid the influence of the photon's polarization on the electron angular distribution we inspect polarization-averaged molecular-frame photoelectron angular distributions, which have been shown to capture the three-dimensional molecular structures quite well \cite{Williams2012PhysRevLett.108.233002,Menssen_2016,Ota.2021,Yoshikawa.2024,kuraoka2024} and allow for chiral sensing independent of the light polarization \cite{PhysRevA.109.L060802}. They can be displayed in the same polar coordinate system as used in Figs.~\ref{Fig3}(a) and \ref{Fig3}(b) with $\theta_e$ being the angle of the photoelectron to the normal of the molecular plane [see Fig.~\ref{Fig1}(a)] and $\phi_e$ being the respective azimuthal electron emission angle where $\phi_e=90^\circ$ is defined by the carbon momentum. For a planar molecular structure, these polarization-averaged photoelectron angular distributions are, by definition, symmetric upon reflection at the molecular plane, i.e., they depend only on the modulus of $\cos\theta_e$, and not on its sign. A robust measure of the out-of-plane torsion is thus the mean value $\langle \cos\theta_e \rangle$. For a planar molecular structure or an equal amount of R and S enantiomeric forms, i.e., an equal amount of upward and downward torsion structures, one obtains $\langle \cos\theta_e \rangle=0$. Figures~\ref{Fig3}(c) and \ref{Fig3}(d) show the asymmetry $\langle \cos\theta_e \rangle$ as a function of the azimuthal angle $\phi_e$ of the electron in the molecular frame for the two subsets of fragmentation events, as indicated in panels (a) and (b). We observe a clear out-of-plane symmetry breaking ($\langle \cos\theta_e \rangle \neq 0$) of the electron emission. The electrons follow the correlated upward and downward torsion and wagging of the H1 and H2 fragments. The full red lines in Figs. \ref{Fig3}(c) and \ref{Fig3}(d) show our calculated results of the photoelectron angular distributions after taking into account an angular resolution of $\Delta \phi_e =20^{\circ}$. The calculations were carried out in the frozen-core Hartree-Fock approximation using the stationary single center (SC) method \cite{Demekhin134_2011,Demekhin102_2007,Novikovskiy_2022} that allows for an accurate description of angle-resolved molecular photoionization. The calculations show good agreement with the observations. If the symmetry breaking we have observed in the ion fragments would have been the result of a dynamic deformation of an originally planar structure, then we would have found $\langle \cos\theta_e \rangle = 0$ for all $\phi_e$. Thus, the observed out-of-the-plane emission of the electrons, i.e., $\langle \cos\theta_e \rangle \neq 0$, unambiguously demonstrates that our experiment senses the instantaneous chirality of individual ground-state molecules. This chirality is present already at $t=0$ when the photon is absorbed and does not emerge only later during the dissociation.  

To conclude, we have shown that upon measurement, formic acid in its planar vibronic ground state is on the single-molecule level found to ``consist'' of chiral structures, of R or S enantiomers.  Formic acid molecules (as well as most other prochiral planar species) are instantaneously chiral in their ground state, after the photoabsorption. Two consecutive measurements of the same molecule find the same handedness of the molecule.
While this may sound contradictory, the effect is a straightforward consequence of the quantum delocalization of the nuclear wavefunction. In one dimension, this manifests in the often close-to-Gaussian distribution of bond lengths \cite{Nature431Weber,PhysRevLett.108.073202} and the spontaneous breaking of the symmetry in photoionization of the CO$_2$ molecule \cite{PhysRevA.80.032506,PhysRevLett.101.083001}.
In two dimensions, it leads to the fact that linear triatomic molecules such as CO$_2$ extrude into two dimensions and are bent at the single-molecule level \cite{JENSEN2020128087}.
In three dimensions, it leads, as we have shown, to a three-dimensionality and in prochiral cases can trigger instantaneous chirality and spontaneous symmetry breaking of planar molecules upon measurement.

\begin{acknowledgments}
This work was funded by the Deutsche Forschungsgemeinschaft (DFG, German Research Foundation) -- Project No. 328961117 -- SFB 1319 ELCH (Extreme light for sensing and driving molecular chirality). The experimental setup was supported by BMBF. The calculations were supported, in part, through the Maxwell computational resources operated at DESY. F.T. acknowledges funding by the DFG -- Project 509471550, Emmy Noether Programme. {J.B.W. acknowledges support from the National Sciences Foundation under Award No. NSF-2208017.} We acknowledge DESY (Hamburg, Germany), a member of the Helmholtz Association HGF, for the provision of experimental facilities. Parts of this research were carried out at PETRA~III and we would like to thank M.~Hoesch and his team for assistance in using beamline P04. Beamtime was allocated for proposal H-20010092. We are grateful to Jochen Mikosch for bringing the work on CO$_2$ to our attention.
\end{acknowledgments}

\bibliographystyle{apsrev4-2}
\bibliography{main_bib}

\end{document}